\documentclass[astrosymb,twocolumn]{aastex701}
\usepackage{graphicx}
\usepackage{amsmath}
\usepackage{amssymb}
\usepackage{bm}
\usepackage{hyperref}
\hypersetup{
  colorlinks   = true,
  citecolor   = blue
}
\usepackage{natbib}
\usepackage{color}
\usepackage{makecell}
\usepackage{soul} % st

\def\apjs{ApJS}
\def\aap{A \& Ap}

\def\apjl{ApJL}
\def\aap{A\&A}

\newcommand{\beq}{\begin{equation}}
\newcommand{\eeq}{\end{equation}}
\def\bar{\overline}
\newcommand{\abra}[1]{\left\langle{#1}\right\rangle}

\newcommand{\del}{{\bm\nabla}}

\def\bmu{\bm{u}}

\newcommand{\ReN}{\text{Re}}

\newcommand{\wT}{\omega_\text{T}}
\newcommand{\wc}{\omega_\text{c}}
\newcommand{\wD}{\omega_\text{d}}
\newcommand{\nut}{\nu_\text{t}}

\newcommand{\Ra}{\text{Ra}}
\newcommand{\cv}{c_\text{v}}
\newcommand{\cp}{c_\text{p}}

\newcommand{\urms}{u_\text{c}}
\newcommand{\Co}{\text{Co}}

\newcommand{\cs}{c_\text{s}}

\newcommand{\Ma}{\text{Ma}}
\newcommand{\wo}{\omega_\text{o}}
\newcommand{\wf}{\omega_\text{f}}
\newcommand{\uc}{u_\text{c}}
\newcommand{\uT}{u_\text{T}}

\newcommand{\expi}[1]{\text{e}^{-\ii{#1}}}
\newcommand{\ii}{\text{i}}

\newcommand{\tw}{\tilde\omega}
\newcommand{\tf}{\tilde f_0}
\newcommand{\twD}{{\tilde\omega}_\text{d}}

\begin{document}

\title{Efficiency of Tidal Dissipation in Convective Flow Under Rapid Tidal Forcing}

\author[0000-0002-2991-5306]{Hongzhe Zhou}
\email[show]{hzzhou@sspu.edu.cn}
\affiliation{Department of Physics, School of Mathematics, Physics and Statistics, Shanghai Polytechnic University, 2360 Jin Hai Road, Shanghai, 201209, China}
\affiliation{Tsung-Dao Lee Institute, Shanghai Jiao Tong University, Shanghai, 201210, China}

\author[0000-0002-1934-6250]{Dong Lai}
\email[show]{dong.lai@sjtu.edu.cn}
\affiliation{Tsung-Dao Lee Institute, Shanghai Jiao Tong University, Shanghai, 201210, China}
\affiliation{Department of Astronomy and Cornell Center for Astrophysics and Planetary Science, Cornell University, Ithaca, NY 14853, USA}

\begin{abstract}
For close binaries and star-planet systems, tidal interactions mediate the energy transfer between the orbital motion and the internal flows of the bodies involved, thus playing a central role in their evolution. For equilibrium tides, the associated energy transfer is commonly modeled through an effective viscosity acting on the tidal flow. However, the scaling of viscous dissipation efficiency with tidal frequency $\omega_\text{T}$ remains debated, particularly when $\omega_\text{T}$ greatly exceeds the convective eddy turnover frequency $\omega_\text{c}$. Previous numerical studies have addressed this issue by subjecting a turbulent convective flow to an oscillating background shear mimicking equilibrium tides. In this work, we adopt a novel three-layered convective box --- designed to represent a stellar convection zone sandwiched between two stable layers --- driven by an external periodic forcing. We quantify tidal dissipation efficiency by the forcing power on the flow in steady state. Our results yield a shallower scaling of tidal power per unit mass with $\omega_\text{T}$ than reported in earlier shear-flow simulations. This scaling is consistent with the prediction by \cite{Terquem2021}, suggesting that the effective turbulent viscosity depends only weakly on $\omega_\text{T}$, although our simulations are restricted to $\omega_\text{T}\lesssim 10\omega_\text{c}$. Moreover, we find no evidence of inverse energy transfer (or ``negative viscosity''), a phenomenon observed in some prior shear-flow simulations. We further investigate the influence of rotation within the same local framework. Slow rotation ($\Omega\lesssim \omega_\text{T}$) tends to enhance the tidal power, whereas fast rotation ($\Omega\gtrsim\omega_\text{T}$) significantly suppresses it. We discuss the limitations of our approach and the broader implications of our findings.
\end{abstract}

%\keywords{}

\section{Introduction}
\label{sec:introduction}

Tidal interactions play important roles in the evolution of binary stars, planetary, and planet-moon systems. 
The dissipation of the orbital energy through tides directly influence various observable phenomena, such as the circularization of binary stars, orbital decay of hot Jupiters, tidal spin-up of planet-host stars, and spin-orbit alignment in binary and exoplanet systems
\citep[e.g.,][]{NorthZahn2003,MeibomMathieu2005,Albrecht+2012,Ogilvie2014, Price-WhelanGoodman2018,Beck+2018,Penev+2018,PatelPenev2022,Bashi2023,Ilic+2024,Barker2025}.

The dissipation of equilibrium tides (forced f-mode oscillations) in
stellar convection zone has long been a subject of discussion and
debate.  Starting with the pioneering work of \cite{Zahn1966}, it is
commonly assumed that the convective flow has an effective turbulent
eddy viscosity, given by $\nut\simeq\urms h_0/3$, where $\urms$ is the
root-mean-square (rms) velocity of the convective turbulence, and
$h_0$ is the pressure scale height of the convection zone.  However,
when the tidal frequency, $\wT$, exceeds the turnover rate of the
convective turbulence, $\wc$, or when the ratio
\beq
\tw\equiv\frac{\wT}{\wc}
\eeq
exceeds unity, it has been speculated
that tidal dissipation will be suppressed.  \cite{Zahn1966} argued
that the correlation time of the energy-dominant eddies should be
replaced by $\wT^{-1}/2$ when $\tw>1$, so that the suppression is
linear, $\nut\propto \tw^{-1}$.  On the other hand, if one assumes
that the energy transfer is primarily mediated by the eddies whose
turnover rate is comparable to the tidal frequency (i.e., the resonant
eddies), the turbulent viscosity associated with these eddies will be
$\nut\propto \tw^{-2}$ for a Kolmogorov turbulence \citep[][but see
also Appendix~\ref{appx:w-2} for an argument without invoking the
resonant-eddy assumption]{GoldreichNicholson1977,OgilvieLesur2012}.
Indeed, such a scaling was found in several numerical simulations of Rayleigh-B\'enard convection with the tidal flow imposed through boundary conditions
\citep{OgilvieLesur2012, Duguid+2020a,
  Duguid+2020b}.  Using data from more realistic stellar convection
zone simulations, \cite{Penev+2009} obtained a shallower scaling of
$\nut\propto \tw^{-1}$ due to the difference in the kinetic energy
spectrum.  Global simulations of spherical convection have also revealed
a transitional regime where $\nut\propto \tw^{-1}$ when
$1<\tw\lesssim 5$, and $\nut\propto \tw^{-2}$ when $\tw\gtrsim 5$
\citep{VidalBarker2020a, VidalBarker2020b}.

Alternatively, \cite{GoodmanOh1997} argued that $\nut$ should be
associated with the frequency spectrum of the turbulent kinetic energy,
$\hat E(\omega)$, yielding $\nut\propto \tw^{-2}$ or $\tw^{-5/3}$ depending on  how the eddy correlation time scales with the wavenumber.  This
connection to $\hat E(\omega)$ has been verified by \cite{Duguid+2020b}, who found an intermediate scaling $\tw^{-0.5}$ in the range $10^{-2}\leq \tw\leq10^{-0.5}$ that is likely caused by a corresponding slope in $\hat E(\omega)$.

More recently,
\cite{Terquem2021,Terquem2023}
argued that, in the $\tw\gg1$ regime, the mean Reynolds stress that
mediates the energy transfer is provided by the more rapid tidal flow,
rather than the convective turbulence as in the conventional picture.
Dissipation then arises from correlations between the oscillatory velocity
fluctuations and the background convective flow gradient.
This results in a much higher tidal dissipation rate which
does not depend on the tidal frequency and could explain a number of
observations \citep{TerquemMartin2021}.  However, robust numerical
evidence supporting the argument of \cite{Terquem2021,Terquem2023} remains to be
demonstrated.
Most of the existing simulations, such as those mentioned above, as well as
\cite{BarkerAstoul2021}, which attempted to test the \cite{Terquem2021}
theory, adopted an impenetrable vertical boundary condition; it has been argued
that such simulation setup does not reflect realistic systems and may
lead to inconsistency with \cite{Terquem2021,Terquem2023}.

In this work, we aim to determine the level of suppression for the
tidal dissipation when $\tw \gtrsim 1$, using simulations where (i) turbulent
convection is set up in a three-layer model that allows for
overshooting, and (ii) tides are introduced by an external periodic
forcing.  Our simulation setups are explained in
Section~\ref{sec:numerics} for the turbulent convection, and in
Section~\ref{sec:forcing} for the tidal forcing.
We then present and analyze the numerical results in Section~\ref{sec:results}.
We conclude in Section~\ref{sec:conclusion}.
During the final preparation of this manuscript, we noticed the work
of \cite{Terquem+2026} who have adopted a similar numerical setup; 
we discuss the similarities and differences between our work and theirs.

%%%%%%%%%%%%%%%%
\section{Numerical Setup of Turbulent Convection}
\label{sec:numerics}

Previous numerical studies of tidal dissipation in convective flows
have employed the Rayleigh-B\'enard convection setup
\citep[e.g.,][]{OgilvieLesur2012,Duguid+2020a}.
The spatial distribution of the turbulent viscosity was found to be enhanced
near the top and the bottom boundaries \citep{Duguid+2020a}, which could be
an artifact resulting from the impenetrable walls.  In this work, we
follow \cite{Brandenburg+1996} to set up a three-layer convective
turbulence with top and bottom overshooting layers, to alleviate the
potential influence of the vertical boundaries.  We also relax the
Boussinesq approximation and solve the equations for compressible
hydrodynamics, although the effect of compressibility is expected to
be small in stellar convection zones.

We use the publicly available \textsc{Pencil Code} \citep{JOSS2021}.
The simulation domain is a Cartesian box with dimensions
$[-L/2,L/2]^2\times[-0.68L,1.32L]$, representing a local patch in the
northern hemisphere on the primary star.  The center of the box is
located at spherical coordinates $(R_0,\theta_0,\phi_0)$, and the
$\hat{\bm x}$, $\hat{\bm y}$ and $\hat{\bm z}$ directions of the box
are aligned with the polar ($\hat{\bm \theta}$), the azimuthal
($\hat{\bm \phi}$) and the radial ($\hat{\bm R}$) directions,
respectively.  We solve the equations of compressible hydrodynamics in
the co-rotating frame of the star,
\begin{align}
\partial_t\rho+\del\cdot\left(\rho\bmu\right)=&0,\\
\partial_t\bmu+\bmu\cdot\del\bmu=&-\frac{1}{\rho}\del p-
2\bm\Omega\times\bmu+\frac{1}{\rho}\del\cdot\left(2\nu\rho{\bf\mathcal{S}}\right)\notag\\
&+\bm g+\bm f_\text{T},
\label{eqn:dudt}\\
T\left(\partial_t s+\bmu\cdot\del s\right)=&
\frac{1}{\rho}\del\cdot\left(\kappa\del T\right)
+2\nu \mathcal{S}^2+\mathcal{C},
\label{eqn:dsdt}
\end{align}
where $\rho$, $\bmu$, $p$, $T$ and $s$ are the density, velocity, pressure, temperature and entropy, respectively;
$\bm\Omega=\Omega(-\sin\theta_0,0,\cos\theta_0)$ is the rotation rate of the star;
$\bm g=-g_0\hat{\bm z}$ is the gravity of the star with $g_0=GM/R_0^2$,
and $G$ and $M$ are the gravitational constant and the mass of the star, respectively;
$\mathcal{S}_{ij}=\left(\partial_i u_j+\partial_j u_i-2\delta_{ij}\partial_ku_k/3\right)/2$ is traceless rate-of-strain tensor;
$\nu$ is a uniform viscosity,
and $\kappa$ is a non-uniform heat conductivity to be specified below;
$\bm f_\text{T}$ is the tidal forcing to be specified in the next section.
We take the ideal-gas equation of state,
\beq
p=(\gamma-1)\cv\rho T,
\eeq
where $\gamma=\cp/\cv=5/3$ and $\cp$ and $\cv$ are specific heats at constant pressure and volume, respectively.
The function $\mathcal{C}$ in Equation~(\ref{eqn:dsdt}) is a Newtonian cooling term applied near the top boundary of the box,
\beq
\mathcal{C}(t,z)=-\cp\frac{T(t,z)-T_0}{\tau_\text{cool}}f_\text{cool}(z),
\eeq
where $T_0$ is the initial temperature at the top boundary,
$\tau_\text{cool}= 8.66 L/c_\text{s,0}$ is the cooling time scale,
$c_{\text{s},0}^2=(\gamma-1)\cv T_0$ is the initial sound speed at the
top layer, and
$f_\text{cool}=\text{exp}\left[-25(z-z_\text{top})^2/2L^2\right]$ with
$z_\text{top}=1.32L$ being the location of the top boundary.  In the
horizontal directions we employ the periodic boundary condition.

The computational domain consists of a convectively unstable layer in
the region $z_0\leq z\leq z_0+L$ and two stable overshoot layers at
$z<z_0$ and $z>z_0+L$, where $z_0=0$.  The convective stability is
controlled by the values of $\kappa$ and vertical temperature gradient
(essentially the Rayleigh number) in the three regions.
The bottom boundary has a constant heat flux injection, while the top
boundary is approximately hold at a constant temperature given by the
cooling term $\mathcal{C}$.  As such, even though both the top and the
bottom boundaries are impenetrable and stress-free, they have a much weaker effect on the
convection compared to the previous one-zone setups.  A radiating top
surface was mentioned by \cite{Terquem2023} to be necessary to carry
away the kinetic energy due to tidal forcing.  In \cite{Terquem+2026},
a similar radiative cooling and free top surface is introduced and
argued to be crucial to produce a finite tidal work.

To specify the initial condition, we consider a reference equilibrium
state that is homogeneous in the horizontal directions and at
hydrostatic equilibrium in the vertical direction.  The heat
conductivity is chosen to be constant in time but non-uniform in $z$ to control the
stability of each layer: We use
$\kappa=\kappa_0\times\left\{1,2,4\right\}$ in the three regions from
top to bottom, respectively, connected by smooth step functions, with
$\kappa_0$ being a constant.  The initial heat flux is uniform and
chosen to be $F=\kappa_0 g_0/[(\gamma-1)\cv]$, so that the initial
temperature gradient is
$\partial_z T(t=0)=-F/\kappa(z)$. The initial
temperature at the top surface then determines the initial $T(z)$,
which in turn gives the initial density profile through the
hydrostatic equilibrium.  We also add small velocity perturbations
to the initial equilibrium state.

For reference, we define the Reynolds number and the Coriolis number by
\beq
\ReN=\frac{\urms {h_0}}{\nu},\quad
\Co=\frac{2\Omega {h_0}}{\urms},
\label{eqn:ReN_Co}
\eeq
where
$\urms$ is the root-mean-squared velocity in the steady state,
and $h_0=c_\text{s,0}^2/g_0$ is the pressure scale height.
We denote $\wc=\urms/h_0$ as the convective frequency.
We also define the mean density in the computational domain as $\bar\rho$,
and $\bar\chi=\kappa(z_0)/\gamma\bar\rho$.
Hence the Rayleigh number in the unstable layer and the Prandtl number are defined as
\beq
\Ra=\frac{g_0L^4}{\nu\bar\chi}\left.\frac{\partial\ln T}{\partial z}\right|_{z=z_0},\quad
\Pr=\nu/\bar\chi,
\eeq
respectively.

%%%%%%%%%%%%%%%%%%%%%%%%%%%%%%%
\section{Tidal Forcing Prescription and Diagnostics}
\label{sec:forcing}

Equilibrium tides have scales of order of the stellar radius, which can be
much larger than the turbulence scale of the convection zone.
For local-box simulations, the tidal flow can be well approximated by
a linear shear, and a common practice is to impose it as a background
flow \citep{OgilvieLesur2012, Duguid+2020a, Duguid+2020b} using the
shearing-box setup.  Similar method has also been used in global
spherical convection simulations with Boussinesq approximation
\citep{VidalBarker2020a, VidalBarker2020b}, where the tidal flow is
prescribed and only the turbulent deviation from the tidal flow is
simulated.

In this work, we introduce the tidal effect in a different way, by 
by adding a non-uniform body force in our simulation Cartesian box.
We derive the force by considering the lowest-order non-uniform part
of the binary gravitational force, and modify it to accommodate
the periodic boundary condition in the horizontal directions.
Our tidal force is physically motivated,
although it shares both the essential periodic and shearing natures with that used by
\cite{Terquem+2026}.

\subsection{Derivation of Tidal Forcing}

The form of the tidal forcing is derived by linearizing the gravity
from the secondary at the location of the simulation box on the
primary star.

Consider a circular binary with semi-major axis $a$ and orbital
frequency $\wo$. In the co-rotating frame of the primary, at the
position $\bm R=(R,\theta,\phi)$ (in spherical coordinates), the
time-dependent quadrupole tidal potential is
\beq
\Phi=-\frac{3Gm R^2}{4a^3}\sin^2\theta\cos(\wT t-2\phi),
\eeq
where $m$ is the companion mass, 
$\wT=2(\omega_\text{o}-\Omega)$ is the tidal frequency, and $\Omega$ is the 
frequency of rotation of the primary (assumed to the aligned with the orbit).
The simulation domain represents a small region in the vicinity of
$\bm R_0=(R_0,\theta_0,\phi_0)$.  We denote the position vector in the
simulation box as $\bm x$.  Expanding $\Phi$ around $\bm R_0$ using
small $\delta\theta=x/R_0$ and $\delta\phi=y/(R_0\sin\theta_0)$, we find
\begin{align}
\Phi(\bm R_0+\bm x)=
&\Phi(\bm R_0)+x_j\partial_j \Phi(\bm R_0)+\frac{1}{2}x_jx_k\partial_{jk}\Phi(\bm R_0)\notag\\
&+\mathcal{O}\left(x^3/R_0^3\right).
\end{align}
On the right-hand side, the first term is uniform in the box, and the second term
induces a small uniform body force which we ignore.
The third term gives the tidal gravity of interest,
\beq
g_i=-x_j\partial_{ij}\Phi(\bm R)=\frac{3Gm}{2a^3}h_{ij}x_j,
\label{eqn:gT}
\eeq
where
\beq
{\bf h}=\begin{pmatrix}
\cos2\theta_0\cos\wT t & 2\cos\theta_0\sin\wT t & \sin2\theta_0\cos\wT t \\
2\cos\theta_0\sin\wT t& -2\cos\wT t & 2\sin\theta_0\sin\wT t \\
\sin2\theta_0\cos\wT t & 2\sin\theta_0\sin\wT t & \sin^2\theta_0\cos\wT t
\end{pmatrix},
\eeq
and we have taken $\phi_0=0$.%
\footnote{%
Note that $\bm g$ is curl-free but not divergence-free, although $\del\Phi$ is solenoidal.
Nevertheless, the divergence of $\bm g$ is small and the flows in our simulations always remain nearly incompressible.}

Although Equation~(\ref{eqn:gT}) is the desired tidal force,
we shall adopt a modified form to accommodate the periodic boundary conditions,
\beq
\bm f_\text{T}=f_0 g_0{\bf h}\cdot\left(
\hat{\bm x}\sin\frac{2\pi x}{L}+\hat{\bm y}\sin\frac{2\pi y}{L}
+\hat{\bm z}\frac{z-z_\text{mid}}{L}
\right),
\label{eqn:fT}
\eeq
where the linear dependences on $x$ are $y$ in Equation~(\ref{eqn:gT})
are replaced by sinusoidal functions,
and the dependence on $z$ is shifted to center at the
middle of the convective layer, $z_\text{mid}$, to ensure zero net
force over the whole simulation volume.  We have also introduced the
dimensionless forcing amplitude
\beq
f_0=\frac{3}{2}\frac{L}{R_0}\frac{m}{M}\left(\frac{R_0}{a}\right)^3.
\label{eqn:f0_fid}
\eeq
For example, with $L/R=0.3$, $m/M=m_\text{J}/M_\odot$, 
$R_0=R_\odot$, and $\wT=2\pi\ \text{d}^{-1}$, one finds $f_0\sim 10^{-8}$.
Such a value is
smaller than the microscopic viscous dissipation term, which 
is limited by the numerical resolution.
The achievable values of $f_0$ must be artificially increased to
bypass the viscosity limitation, but still in the linear-tide regime,
the latter being quantified by the tidal displacement.
Note that in our simulation setup, the magnitude of the tidal displacement
is $\simeq f_0g_0\wT^{-2}$ (see below), rather than $f_0g_0\wf^{-2}$ as in real stars (recall that $g_0=GM/R_0^2$).
The ratio between the tidal displacement and the radius of the star is thus
\beq
\tf\equiv \frac{f_0g_0}{\wT^2 R_0}.
\eeq
Hence we properly choose $f_0$ to make $\bm f_\text{T}$ larger than the viscous term,
but keep $\tf\ll 1$ in all of our simulations.

\begin{figure*}
\centering
\includegraphics[width=0.9\textwidth]{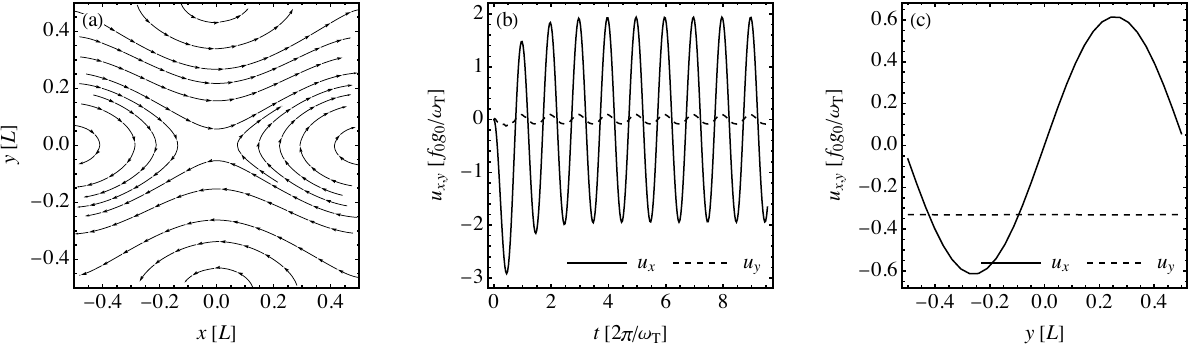}
\caption{Flow visualization to validate our tidal forcing presciption with no turbulence.
(a) Streamlines of the horizontal velocity fields in the $z=0$ plane at $t=9.55\times 2\pi/\wT$.
(b) Time variation of the horizontal velocity fields at $(x,y,z)=(0,L/4,0)$.
(c) Spatial variation of the horizontal velocity fields at $t=9.26\times 2\pi/\wT$ and $x=z=0$.}
\label{fig:fT_snap}
\end{figure*}

We can show that our tidal forcing prescription realizes
a periodic shear flow using an simplified setup: A homogeneous fluid is forced by $\bm
f_\text{T}$ with $\theta_0=0$ in a three-dimensional periodic box,
with no background turbulence, stratification or gravity.  A snapshot
of the horizontal flow in the $z=0$ plane, along with the temporal and
spatial variations of the horizontal velocity, is shown in
Figure~\ref{fig:fT_snap}.  The velocity fields are indeed periodic in
both space and time, and the magnitude is of order $f_0g_0\wT^{-1}$.

For our convective turbulence simulations, we set $\bm f_\text{T}=\bm
0$ in the top and the bottom convectively stable zones to ensure that
tidal flows are only driven in the interested middle layer.  It also
ensures that the flows in the top and bottom layers do not contribute
to the tidal power (see Section~\ref{sec:diag}).

%%%%%%%%%%%%%%%%%%%%
\subsection{Critical frequencies}
\label{sec:crit_freq}

We now consider the dimensionless tidal frequency $\tw\equiv\wT/\wc$
that can be achieved in our simulations.  Equating the tidal frequency to the
sound-crossing time in the simulation box, we obtain a critical value of
\beq
\tw_\text{s}=\frac{\cs/L}{\wc}
=\frac{\cs {h_0}}{L\urms}=\Ma^{-1}\frac{{h_0}}{L},
\eeq
where $\Ma=\urms/\cs$ is the Mach number.  For all of our simulations,
we have ${h_0}/L\simeq0.2$ and $\Ma\simeq0.02$, yielding a critical
value $\tw_\text{s}\simeq10$.

The upper bound of meaningful values of $\tw$ is also constrained by
the largest eddy turnover rate, which applies to the eddies at the
Kolmogorov scale $l_\nu$.  With a $-5/3$ kinetic energy spectrum, the
turnover rate at $l_\nu$ is $\urms{h_0}^{-1}(l_\nu/{h_0})^{-2/3}$,
yielding
\beq
\tw_\nu=\frac{\urms{h_0}^{-1}(l_\nu/{h_0})^{-2/3}}{\urms/{h_0}}
=\ReN^{1/2}.
\label{eqn:w_bound}
\eeq
Beyond this value, the tidal frequency is larger than the turnover rate of
smallest eddies in the flow.

Astrophysically relevant tidal frequencies satisfy $\tw\ll
\tw_\text{s} \ll \tw_\nu$.  In our simulations, we typically have
$\ReN\simeq100$ and hence the limits set by $\tw_\text{s}$ and
$\tw_\nu$ are similar.  We observe that, beyond this limit,
the results deviate substantially from the scaling of the data below
it (see Figure~\ref{fig:H_epsilon_w}).

\subsection{Diagnostic quantities and interpretation}
\label{sec:diag}

\begin{figure*}
\centering
\includegraphics[width=0.8\textwidth]{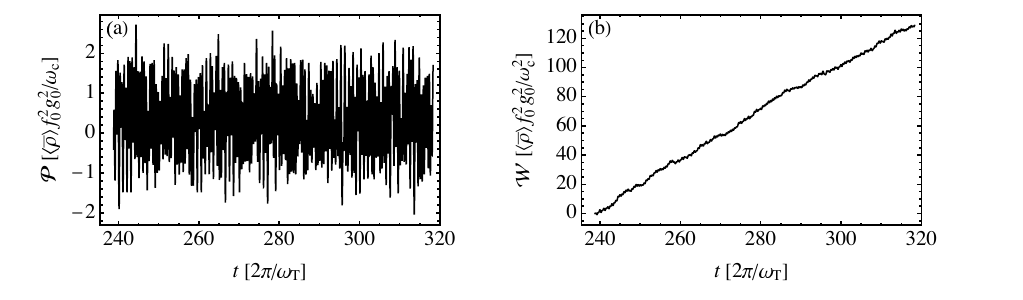}
\caption{Time series of the tidal power and the work done in run~\texttt{A1}.}
\label{fig:w1_W_vs_t}
\end{figure*}

The main diagnostic quantity in our simulations is the mean tidal
power in the convective region,
\beq
\mathcal{P}(t)=\abra{\rho\bm u\cdot\bm f_\text{T}},
\eeq
where $\abra{\cdot}$ denotes volume average in
a statistically steady turbulent state (note that $\bm f_\text{T}$
vanishes outside the convective layer $z_0\leq z\leq z_0+L$).  To find
the average tidal power $\bar{\mathcal{P}}$ over long time scales, we
calculate the work done up to some time $t$, $\mathcal{W}(t)=\int_0^t
\mathcal{P}(t')\ \text{d}t'$, and fit it with a linear function whose
slope is identified as $\bar{\mathcal{P}}$.  The dimensionless mean
tidal power per unit mass is defined as
\beq
\hat{\mathcal{P}}=\frac{\bar{\mathcal{P}}}{\abra{\bar\rho} f_0^2g_0^2\wc^{-1}}.
\label{eqn:hatP}
\eeq
In Figure~\ref{fig:w1_W_vs_t}, we show a sample of curves
$\mathcal{P}$ and $\mathcal{W}$.  Although $\mathcal{P}$ fluctuates
significantly in time, it results in a net work done and therefore a
well-defined average power $\bar{\mathcal{P}}$.  The full set of
curves for all of our simulation runs are shown in Appendix~\ref{appx:W}.

All previous simulations and analysis, including those of
\cite{Terquem+2026}, have calculated the tidal dissipation through the
coupling between the Reynolds stresses of mean and turbulent flows.
Here we have taken a different approach by considering the work done
by the tidal force, which directly reflects the dissipation rate of
tidal potential.  It is then necessary to derive the expected scaling
of $\hat{\mathcal{P}}$ under different assumptions about the tidal dissipation rate.

For a solar-type star with a convective envelope, we can schematically write
the equation of motion for the (large-scale) tidal displacement $\bm\xi$ as
\beq
\ddot{\bm\xi}+\wf^2\bm\xi+\bm\Gamma=\bm F_0\expi{\wT t},
\label{eqn:harmonic_oscillator}
\eeq
where $\omega_\text{f}$ is the f-mode frequency,
$\bm F_0$ (of order $f_0g_0$; see Equation~(\ref{eqn:fT})) is the
space-dependent amplitude of the tidal forcing, and $\bm\Gamma$
includes the microscopic viscous damping and the coupling between the
large-scale tidal flow and the small-scale turbulent flow.
If we assume that $\bm\Gamma$ can be written as a damping term,
$\bm\Gamma=\wD\dot{\bm\xi}$, with some damping rate $\wD$, then
Equation~(\ref{eqn:harmonic_oscillator}) can be solved for the steady-state:
\beq
\bm\xi(t)=\frac{\bm F_0\expi{\wT t}}{\wf^2-\wT^2-\ii\wD\wT}.
\label{eqn:xi(t)}
\eeq
We shall now distinguish two scenarios based on the presence or absence of the f-mode:
(i) for a realistic star, we have $\wf\gg\wT$, and
(ii) for a local-box simulation, we do not have global normal modes and thus $\wf=0$. The mean tidal power per unit mass is therefore different
in the two cases, namely,
\beq
\bar{\mathcal{P}}=
\Re\left(\overline{\dot{\bm\xi}^*\cdot \bm F_0\expi{\wT t}}\right)
\simeq\left\{\begin{aligned}
&\frac{\wD\wT^2 f_0^2g_0^2}{\wf^4}, & ~~~\wf\gg\wT \\
&\frac{\wD f_0^2g_0^2}{\wT^{2}+\wD^2}, & \wf=0
\end{aligned}\right.,
\label{eqn:re_xi_f}
\eeq
where the overline indicates a time average and we have identified $F_0=f_0g_0$.
Note that the spatial dependence of $\bm f_\text{T}$ (see Equation~(\ref{eqn:fT}))
may introduce an extra factor of order unity, but our focus here is
on how the tidal power scales with $\wD$, which is not affected.
In particular, the $\wf=0$ branch of Equation~(\ref{eqn:re_xi_f}) is relevant
for the simulations presented this work.

When convective turbulence is present, we expect that the energy
transfer rate between the tidal flow and the turbulence is much larger
than the viscous damping rate of the tidal flow, and therefore
dominates the $\Gamma$ term
in Equation~(\ref{eqn:harmonic_oscillator}). In this case, the appropriate functional
form $\wD(\tw)$ is under debate, particularly regarding how it is
suppressed when $\tw\gg1$.

We consider the ansatz
\beq
\wD=C(\tw)\wc \times\left\{\begin{aligned}
& \frac{1}{1+\tw^p}, & \text{if } p>0 \\
& 1, & \text{if }p=0
\end{aligned}\right..
\label{eqn:wD}
\eeq
Note that the prefactor $C$ may also depend on $\tw$.  In the
$\tw\ll 1$ regime, all previous studies have indicated
that $\wD$ is equal to the turbulent diffusion rate, given by
$\wD=\nu_\text{t}k_1^2$ with $\nu_{\rm t}=u_ch_0/3$ and
$k_1=2\pi/L$. Thus
\beq
C(\tw \ll 1)=C_0
=\frac{1}{3}\left(\frac{2\pi h_0}{L}\right)^2,
\label{eqn:C0}
\eeq
which is $\simeq 0.53$ given $h_0/L=0.2$ in our numerical setup.
In the $\tw\gg1$ regime, most previous theories predict finite values of $p$
(see the last column in Table~\ref{tab:wD}).
In contrast, \cite{Terquem2021} argues that the tidal energy dissipation rate
should be given by $D_R=\abra{u'_iu'_j}\partial_i V_j$, where $\bm u'$
is always the faster flow and therefore is the tidal velocity in this
case, and $\bm V$ is the slower convective flow.  The corresponding
dissipation frequency is
\beq
\wD=\frac{D_R}{\uT^2}\simeq \frac{\uT^2\uc/h_0}{\uT^2}=\wc,
\eeq
which leads to $p=0$ and $C(\tw\gg1)=C_\infty=1$.
Hence for the \cite{Terquem2021} prescription, $\wD$ takes the lower branch in
Equation~(\ref{eqn:wD}) with $C$ varying
by a factor of order unity as $\tw$ increases from $\ll 1$ to $\gg 1$;
see the last row in Table~\ref{tab:wD}.

To summarize, for our local-box simulations, the tidal power per unit mass is given
by the lower branch of Equation~(\ref{eqn:re_xi_f}). In units of $(f_0g_0)^2/\wc$ (see 
Equation~(\ref{eqn:hatP})), the tidal power is 
\beq
\hat{\mathcal{P}}(\tw)\simeq \frac{\wD\omega_{\rm c}}{\wT^2
  +\wD^2}
=\frac{\twD}{\tw^2+\twD^2},
\label{eqn:epsilon_th}
\eeq
where $\tilde\omega_\text{d}=\wD/\wc$ depends on $\tilde\omega$.
Together with Equation~(\ref{eqn:wD}), the asymptotic behaviors are
\beq
\hat{\mathcal{P}}(\tw)\to
\left\{\begin{aligned}
&C_0^{-1}, &\tw\ll1,\\
&C_\infty\tw^{-2-p}, &\tw\gg1
\end{aligned}\right.,
\label{eqn:hatP2}
\eeq
for any $p\geq 0$. We aim to determine $\hat{\mathcal{P}}(\tw)$ (and thus
$C(\tw)$ and $p$) from our simulations.

\begin{table*}[htp]
\caption{Summary of the damping rate~(Equation~\ref{eqn:wD}) in different works.}
\begin{center}
\begin{tabular}{lcc}
\hline
\hline
Reference & $C$ & $p$\\
\hline
\cite{Zahn1966} & $C_0$ & $1$ \\
\cite{GoodmanOh1997} & $C_0$ & $2$ or $5/3$ \\
\cite{Penev+2009} & $C_0$ & $1$ \\
\cite{OgilvieLesur2012} & $C_0$ & $2$ \\
\cite{Duguid+2020a,Duguid+2020b} & $C_0$ & $2$ \\
\cite{Terquem2021, Terquem2023} &  \makecell[l]{$\tw\ll1$: $C_0$ \\ $\tw\gg1$: $C_\infty=1$} & $0$ \\
\hline
\end{tabular}
\end{center}
\label{tab:wD}
\end{table*}

\subsection{Limitations of the setup}
\label{sec:limitations}

The middle convective layer in our Cartesian box is meant to represent
a local patch in stellar convection zones.  The limited size of the
box has two consequences.

First, to accommodate the periodic boundaries in the horizontal
direction, the scale of variation of the tidal forcing is reduced from
$\simeq R_0$ to the size of the box $L$ (see Equation~(\ref{eqn:fT})).  We
argue that since the scale separation in our simulations is
$h_0/L\simeq 0.2$, the setup still marginally resembles the scale separation
found in real stars.  Additionally, the sinusoidal dependence of the tidal force
in the horizontal directions makes the forcing vortical.  We argue that
the vortical forcing has a limited influence on the problem, since for
any values of $\theta_0$, $\del\times\bm f_\text{T}$ varies on the
scale of $L$, is periodic in both space and time, and has a zero time
average.  Thus no net vorticity is introduced to the flow over long
time scales.  Furthermore, for the fiducial runs (see the next
section), we have used $\theta_0=0$ for which the volume-averaged
$\del\times\bm f_\text{T}$ vanishes identically.

Second, our simulations differ from the previous ones by how the
tidal flow is realized.  We do not impose a mean flow with constant
period and amplitude; rather, the mean flow is induced by exerting a
force onto the existing convective turbulence.  Hence a perfectly
laminar and periodic component of the flow is not seen.  Rather, a
perturbatively small but coherent deviation from the otherwise
turbulent flow arises as the response of the tidal forcing.  It is
this coherent response that contributes to a finite tidal work.

One might be concerned that by using a local box, our simulation does not produce the correct tidal flow.
For example, the simulation domain does
not host global f-modes, and hence the restoring force
opposing the tidal forcing is missing.  As a result, the tidal
displacement cannot be self-consistently determined; rather, it is 
given by the forcing amplitude and frequency, roughly as $f_0g_0\wT^{-2}$
(see Equations~(\ref{eqn:xi(t)}) and (\ref{eqn:re_xi_f})).  Nevertheless, we have made the tidal displacement
as small as possible and checked that the tidal interaction is in the
linear regime.
In fact, substituting Equation~(\ref{eqn:xi(t)}) into the second term on the
left-hand side of Equation~(\ref{eqn:harmonic_oscillator}) yields
\beq
\ddot{\bm \xi}+\bm\Gamma=
\frac{\wT^2+\text{i}\wD\wT}{\wT^2+\text{i}\wD\wT-\wf^2}
\bm F_0 e^{-\text{i}\wT t},
\eeq
i.e., the effect of the missing f-mode can be effectively incorporated
by lowering the forcing amplitude (since $\wf\gg\wT,\wD$).  The phase
difference between $\bm \xi$ and the forcing is also affected, which
changes the mean power in addition to the lowered amplitude, as 
captured by the lower branch of Equation~(\ref{eqn:re_xi_f}).

To summarize in more physical terms: The tidal power (per unit mass) is
of order $\wT^2|{\bm\xi}|^2/t_\text{d}$, where 
${\bm\xi}$ is the tidal displacement and $t_\text{d}\sim \wD^{-1}$
is the dissipation time. Our simulations are expected to give
$|{\bm\xi}|\sim F_0/\wT^2 \sim f_0g_0/\wT^2$ (assuming $\wT\gg\wD$), and thus
the dimensionless tidal power (in units of $f_0^2g_0^2/\wc$)
is $\hat{\mathcal{P}}\sim \wc\wD/\wT^2$ (see Equation~(\ref{eqn:epsilon_th})).
While our $\bm\xi$
does not match the actual tidal displacement in real stars,
we expect that the tidal
dissipation time $t_\text{d}\sim L^2/\nut$, or the effective viscosity $\nut$,
as determined from computing $\hat{\mathcal{P}}$ numerically,
is not affected by the lack of f-mode in our simulation setup.

%%%%%%%%%%%%%%%%%%%%%%%%%%%%%%%%%%%%%%
\section{Results}
\label{sec:results}

\begin{figure}
\centering
\includegraphics[width=0.5\textwidth]{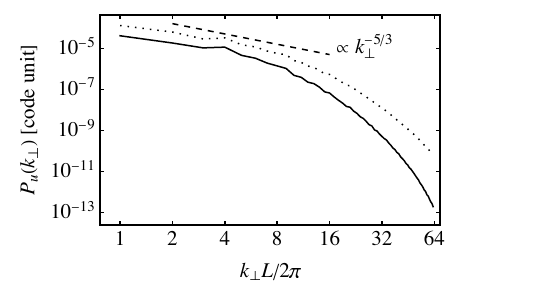}
\caption{The perpendicular spectra of the velocity field for run \texttt{A1},
at the start of the simulation before the tidal forcing is applied (dotted line)
  and at the end of the simulation (solid line).
The dashed line denotes the Kolmogorov scaling $\propto k_\perp^{-5/3}$.}
\label{fig:spectra}
\end{figure}

Our simulation runs are summarized in Tables~\ref{tab:runs} and
\ref{tab:runs_HR} for the non-rotating and rotating cases,
respectively.  The fiducial run parameters are $128\times128\times256$
for the resolution, $\Ra=6.7\times10^8$, $\ReN\simeq100$, and
$\Pr=0.48$ (see Section 2). Each forced run starts from a snapshot of an unforced
counterpart in its steady state, and the unforced rotating runs start
from a snapshot of the unforced non-rotating run.  In
Figure~\ref{fig:spectra}, we plot the perpendicular spectra of the
velocity field, defined as
\beq
P_u(k_\perp)=\int_{-\infty}^{\infty} |\tilde{\bm u}|^2\ \text{d}k_z,\ 
k_\perp=\sqrt{k_x^2+k_y^2},
\eeq
where $\tilde{\bm u}$ is the Fourier transform of ${\bm u}$,
before and after the tidal forcing is applied.
As the forcing is weak, only minor difference is seen and both spectra
are close to the Kolmogorov scaling $\propto k_\perp^{-5/3}$ in the
inertial range.

\begin{table*}[htp]
\caption{Simulation parameters for the non-rotating cases.
Group~\texttt{A} contains the fiducial runs,
while group~\texttt{B} halfvens the forcing amplitude,
and group~\texttt{C} approximately doubles the Rayleigh number and the Reynolds number.
The mean Mach number is $\simeq0.02$ for each run.}
\begin{center}
\begin{tabular}{ccccccc}
\hline
Run & Resolution & $\Ra$ & $\Pr$ & $\ReN$ & $\tw$ & $\tf$\\
\hline
\texttt{A1} & $128^2\times256$ & $6.67\times 10^{8}$ & 0.48 & 103 & $1.29$ & $6.67\times 10^{-3}$\\
\texttt{A2} & $128^2\times256$ & $6.67\times 10^{8}$ & 0.48 & 102 & $2.61$ & $1.67\times 10^{-3}$\\
\texttt{A3} & $128^2\times256$ & $6.67\times 10^{8}$ & 0.48 & 101 & $3.73$ & $8.33\times 10^{-4}$\\
\texttt{A4} & $128^2\times256$ & $6.67\times 10^{8}$ & 0.48 & 102 & $5.23$ & $4.17\times 10^{-4}$\\
\texttt{A5} & $128^2\times256$ & $6.67\times 10^{8}$ & 0.48 & 102 & $7.43$ & $2.08\times 10^{-4}$\\
\texttt{A6} & $128^2\times256$ & $6.67\times 10^{8}$ & 0.48 & 101 & $1.06\times 10^{1}$ & $1.04\times 10^{-4}$\\
\texttt{A7} & $128^2\times256$ & $6.67\times 10^{8}$ & 0.48 & 106 & $1.42\times 10^{1}$ & $5.21\times 10^{-5}$\\
\texttt{A8} & $128^2\times256$ & $6.67\times 10^{8}$ & 0.48 & 102 & $2.08\times 10^{1}$ & $2.60\times 10^{-5}$\\
\texttt{A9} & $128^2\times256$ & $6.67\times 10^{8}$ & 0.48 & 105 & $4.05\times 10^{1}$ & $6.51\times 10^{-6}$\\
\texttt{A10} & $128^2\times256$ & $6.67\times 10^{8}$ & 0.48 & 98 & $6.79\times 10^{-1}$ & $6.67\times 10^{-3}$\\
\texttt{A11} & $128^2\times256$ & $6.66\times 10^{8}$ & 0.48 & 97 & $3.45\times 10^{-1}$ & $6.67\times 10^{-3}$\\
\texttt{A12} & $128^2\times256$ & $6.66\times 10^{8}$ & 0.48 & 97 & $1.72\times 10^{-1}$ & $6.67\times 10^{-3}$\\
\texttt{A13} & $128^2\times256$ & $6.65\times 10^{8}$ & 0.48 & 97 & $8.60\times 10^{-2}$ & $1.07\times 10^{-1}$\\
\texttt{A14} & $128^2\times256$ & $6.65\times 10^{8}$ & 0.48 & 97 & $4.31\times 10^{-2}$ & $2.13\times 10^{-1}$\\
\hline \texttt{B1} & $128^2\times256$ & $6.66\times 10^{8}$ & 0.48 & 97 & $3.45\times 10^{-1}$ & $3.33\times 10^{-3}$\\
\texttt{B2} & $128^2\times256$ & $6.67\times 10^{8}$ & 0.48 & 101 & $5.3$ & $2.08\times 10^{-4}$\\
\texttt{B3} & $128^2\times256$ & $6.67\times 10^{8}$ & 0.48 & 101 & $2.11\times 10^{1}$ & $1.30\times 10^{-5}$\\
\hline \texttt{C1} & $256^2\times512$ & $2.77\times 10^{9}$ & 0.5 & 185 & $1.44$ & $6.67\times 10^{-3}$\\
\texttt{C2} & $256^2\times512$ & $2.77\times 10^{9}$ & 0.5 & 177 & $3.02$ & $1.67\times 10^{-3}$\\
\texttt{C3} & $256^2\times512$ & $2.76\times 10^{9}$ & 0.5 & 179 & $5.97$ & $4.17\times 10^{-4}$\\
\texttt{C4} & $256^2\times512$ & $2.76\times 10^{9}$ & 0.5 & 178 & $1.20\times 10^{1}$ & $1.04\times 10^{-4}$\\
\texttt{C5} & $256^2\times512$ & $2.76\times 10^{9}$ & 0.5 & 178 & $2.40\times 10^{1}$ & $2.60\times 10^{-5}$\\
\hline
\end{tabular}
\end{center}
\label{tab:runs}
\end{table*}

\begin{table}[htp]
\caption{Simulation parameters for the rotating cases.
For all the runs shown here, the resolution is $128^2\times256$,
$\Ra=6.67\times 10^8$, $\Pr=0.48$, and $\tilde f_0=1.67\times10^{-3}$.
The mean Mach number is $\simeq0.02$ for each run.}
\begin{center}
\begin{tabular}{ccccc}
\hline
Run & $\ReN$ & $\Co$ & $\theta_0 [\deg]$ & $\tw$\\
\hline
\texttt{D1} & 98 & 1.63 & 0 & $2.71$\\
\texttt{D2} & 96 & 2.77 & 0 & $2.77$\\
\texttt{D3} & 120 & 5.34 & 0 & $2.23$\\
\texttt{D4} & 121 & 6.82 & 0 & $2.2$\\
\texttt{D5} & 107 & 9.96 & 0 & $2.49$\\
\hline \texttt{E1} & 96 & 1.67 & 22.5 & $2.78$\\
\texttt{E2} & 97 & 2.75 & 22.5 & $2.75$\\
\texttt{E3} & 121 & 5.3 & 22.5 & $2.21$\\
\texttt{E4} & 120 & 6.9 & 22.5 & $2.23$\\
\texttt{E5} & 108 & 9.85 & 22.5 & $2.46$\\
\hline \texttt{F1} & 97 & 1.65 & 45 & $2.75$\\
\texttt{F2} & 99 & 2.69 & 45 & $2.69$\\
\texttt{F3} & 120 & 5.35 & 45 & $2.23$\\
\texttt{F4} & 118 & 6.98 & 45 & $2.25$\\
\texttt{F5} & 98 & 10.8 & 45 & $2.71$\\
\hline \texttt{G1} & 96 & 1.66 & 68 & $2.77$\\
\texttt{G2} & 108 & 2.46 & 68 & $2.46$\\
\texttt{G3} & 126 & 5.07 & 68 & $2.11$\\
\texttt{G4} & 120 & 6.88 & 68 & $2.22$\\
\texttt{G5} & 106 & 10. & 68 & $2.5$\\
\hline \texttt{H1} & 96 & 1.66 & 90 & $2.77$\\
\texttt{H2} & 99 & 3.11 & 90 & $2.71$\\
\texttt{H3} & 121 & 5.29 & 90 & $2.21$\\
\texttt{H4} & 121 & 6.81 & 90 & $2.2$\\
\texttt{H5} & 105 & 10.2 & 90 & $2.55$\\
\hline
\end{tabular}
\end{center}
\label{tab:runs_HR}
\end{table}

\begin{figure*}
\centering
\includegraphics[width=0.8\textwidth]{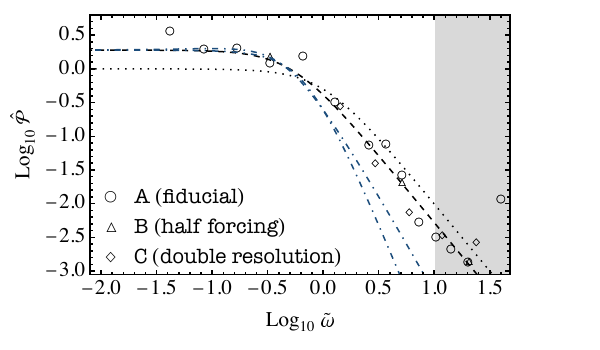}
\caption{The dimensionless tidal energy dissipation rate ($\hat{\mathcal{P}}$; Equation~(\ref{eqn:hatP}))
versus the dimensionless tidal frequency ($\tw=\wT/\wc$) for
non-rotating hydrodynamical runs (group~\texttt{A} in Table 1) and
its variations (groups~\texttt{B} and \texttt{C}).  The shaded area
on the right indicates the $\tw\geq \ReN^{1/2}$ region for
groups~\texttt{A} and \texttt{B}, for which the tidal frequency is
faster than the turnover rate of the smallest eddies in the
simulations.  For group~\texttt{C}, the $\tw\geq \ReN^{1/2}$ boundary
is slightly larger, at $\simeq 10^{1.12}$.
The two blue curves correspond to Equations~(\ref{eqn:wD}) and (\ref{eqn:hatP2}) with $C(\tw)=C_0=0.53$ for $p=1,2$.
The black curves depict the same equations with $p=0$ and
$C(\tw)=C_0$ (dashed) and $C(\tw)=C_\infty=1$ (dotted).
The \cite{Terquem2021} prediction
corresponds to following the black dashed curve in the $\tw\ll1$
regime and the black dotted curve in the $\tw\gg1$ regime.}
\label{fig:H_epsilon_w}
\end{figure*}

In our simulations, group~A contains a set of non-rotating
runs of convective turbulence, and its measured
$\hat{\mathcal{P}}$-$\tw$ relation is presented in
Figure~\ref{fig:H_epsilon_w}.  We also plot four theoretical curves
for reference:
(i) The blue dot-dashed curves are given by
Equation~(\ref{eqn:hatP2}) with Equation~(\ref{eqn:wD}) for $p=1$ and $p=2$, with $C(\tilde\omega)=C_0=(2\pi h_0/L)^2/3\simeq0.53$;
these correspond to the predictions from the turbulent viscosity scenario;
(ii) The black-dashed and the black-dotted curves are
given by Equation~(\ref{eqn:hatP2}) with Equation~(\ref{eqn:wD}) for $p=0$, with
$C(\tilde\omega)=C_0=0.53$ (valid for $\tw\ll 1$, dashed line)
and $C(\tilde\omega)=C_\infty=1$ (valid for $\tw\gg 1$, dotted line);
these correspond to the prediction of \cite{Terquem2021}.
Our numerical data appear to align with the black
curves in the $\tw\in[10^{0},10^{0.7}]$ regime, but the agreement
becomes worse in the $\tw\ll1$ and $\tw\gtrsim10$ regime.  In the
$\tw\in[10,10^{1.5}]$ region, although the data points follow the
slope of the black curves well, we caution that since these
points have entered the $\tw\gtrsim\ReN^{1/2}$ regime (as marked by
the gray area) where the tidal frequency becomes comparable to the
eddy turnover rate at the viscous scale (see
 Section~\ref{sec:crit_freq} and Equation~(\ref{eqn:w_bound})). The
right-most two data points have clearly suffered from the
viscous effects.

We demonstrate the robustness of our fiducial results
(circles in Figure~\ref{fig:H_epsilon_w}) by performing the same calculations
with half the forcing amplitude (triangles), and with
double resolution, half the microscopic viscosity and the heat
conductivity (hence approximately doubling $\ReN$ and $\Ra$; diamonds).
Only minor change in the $\hat{\mathcal{P}}$-$\tw$ relation is seen
for both types of variation; this demonstrates the robustness of our
results and their relevance to astrophysical regimes with even larger
$\ReN$ and lower $f_0$.

We conclude that our results are consistent with $p=0$,
or effectively a frequency-independent turbulent viscosity.
Our results are not consistent with $p=1$ and $p=2$.
However, we note that the agreement between the
data and the black dashed and dotted curves in Figure~\ref{fig:H_epsilon_w} is not decisive
given that the most robust agreement appears in the transition region
where $\tw$ is no more than one decade larger than unity.

\cite{Duguid+2020a} and \cite{Duguid+2020b} have obtained negative
values of $\nut$ when $\tw\gtrsim 10$, which are absent in our
simulations.  We note that the Reynolds numbers in their simulations
are of order $100$ as deduced from the resolution used and the energy
spectra.  Hence the $\tw\gtrsim10$ regime in both works is likely
beyond the realization condition (\ref{eqn:w_bound}).  We also note
that in \cite{Duguid+2020a}, the critical value of $\tw$ beyond which
$\nut$ becomes negative increases with increasing Rayleigh number, which
suggests that such behavior is related to the width of the
inertial range.
It is therefore possible that the negative values of
$\nut$ in their $\tw>10$ regime result from the absence of resonant eddies
interacting with the tidal flow, whereas such resonant eddies are always present in realistic binary systems.

\begin{figure}
\centering
\includegraphics[width=0.5\textwidth]{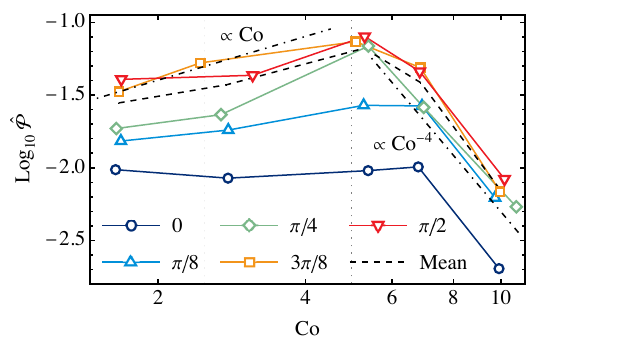}
\caption{The dimensionless tidal energy dissipation rate, $\hat{\mathcal{P}}$, versus the Coriolis number, $\Co$, at $\tw=2.48\pm0.25$ for the rotating
cases at different polar angles (as labeled).
The dashed curve is the
polar-angle-averaged value at each $\Co$.  The left vertical dotted line
indicates $\Omega=\wT/2$, i.e., to the right of which the inertial
waves are excited; the right vertical dotted line indicates
$\Omega=\wT$.  Two reference dot-dashed lines ($\propto\Co$ and
$\propto\Co^{-4}$) are displayed to guide the eye.}
\label{fig:HR}
\end{figure}

Regarding rotating cases, in Figure~\ref{fig:HR} we show the measured
dimensionless tidal power $\hat{\mathcal{P}}$ at $\tw\simeq2.5$ for different values of $\Co$
(defined in Equation~(\ref{eqn:ReN_Co})) and polar angles $\theta_0$.  Note that the tidal forcing also varies
with $\theta_0$ according to Equation~(\ref{eqn:fT}).  The value of 
$\hat{\mathcal{P}}$ depends in a complex way on both
$\theta_0$ and $\Co$: For all the latitudes examined, $\hat{\mathcal{P}}$ is
larger near the equator than at the polar regions, but its variation
with $\Co$ is not uniform.  In particular, near the equator,
$\hat{\mathcal{P}}$ increases with increasing $\Co$ once $\Omega$ is large
enough to
allow for inertial waves to propagate, i.e. $\wT<2\Omega$
\citep[e.g.,][]{Ogilvie2013},
whereas the high-latitude cases only show minor dependence.  At large
rotation rates ($\Co\gtrsim 2\tw$, or equivalently
$\Omega\gtrsim\wT$), the tidal dissipation is suppressed in
all latitudes.
This is likely caused by the suppression of turbulent
transport in fast rotating turbulence \citep{Brandenburg+2009}.

In Figure~\ref{fig:HR}, we also show the polar-angle-averaged value of
$\hat{\mathcal{P}}$ at each fixed $\Co$, defined as
\beq
\bar{\hat{\mathcal{P}}}(\Co)=\int_0^{\pi/2}\hat{\mathcal{P}}(\theta,\Co)\sin\theta\ \text{d}\theta.
\eeq
Overall, a moderate increase in $\bar{\hat{\mathcal{P}}}$ slightly shallower than
$\propto \Co$ is seen in the range $\Omega\lesssim\wT$, and the
suppression at large $\Co$ close to $\propto\Co^{-4}$ persists.  These results suggest
that in our local-box simulations with rotation,
tidal dissipation through turbulent convection remains dominant, and the contribution
to the dissipation through inertial waves is relatively small.
We note that the excitation and dissipation of wave-like modes in a shell
geometry can be quite different from that in a box, especially with respect to the
distance that perturbations can propagate and the sites where they
dissipate.  A similar numerical setup in spherical geometry is
required to fully compare the contributions from the
turbulence-mediated versus the wave-mediated dissipation.

%%%%%%%%%%%%%%%%%%%%%%%%%%%%%%%%%
\section{Summary and Discussion}
\label{sec:conclusion}

\subsection{Key Results}

The efficiency of tidal dissipation in stellar convection zone is a
long-standing problem. In particular, it remains controversial as to
whether and how the tidal dissipation rate is suppressed when the
tidal frequency $\wT$ is much larger than the convective
turnover rate $\wc$. To tackle this problem, we employ a new
hydrodynamical model in which (i) the turbulent convection zone is
sandwiched between two stable layers, and (ii) the equilibrium tidal
flow is driven by an external force rather than being imposed.
The size of our ``local'' simulation box $L$ is intended to represent
a significant portion of the stellar radius (e.g. $L/R\sim 0.3$), and the external
force varies periodically in the $x,~y$ directions (perpendicular to gravity)
to mimic the tidal force.
We quantify tidal dissipation efficiency by the time-averaged forcing
power on the flow in steady state.
Our simulations yield a shallow scaling between the dimensionless tidal power, $\hat{\mathcal{P}}$
(defined by Equation~(\ref{eqn:hatP})),
and the dimensionless tidal frequency, $\tw=\wT/\wc$,
in the $\tw\gg1$ regime (see Figure~\ref{fig:H_epsilon_w}).
Indeed, the tidal power (or tidal dissipation rate) per unit mass $\mathcal{P}$
can be generally written as
\beq
  \mathcal{P}\sim |\xi|^2\wT^2/t_\text{d},
\eeq
where $t_\text{d}=\wD^{-1}$ is the effective damping time of the equilibrium tide, and
$\xi$ is the tidal amplitude (displacement). Since in our numerical model
$\xi\sim F_0/(\wT^2+i\wD\wT)$ (see Equation~(\ref{eqn:xi(t)})), where
$F_0=f_0g_0 \sim GmL/a^3$ is the tidal force across the computation domain $L$,
the dimensionless tidal power (defined by Equation~(\ref{eqn:hatP})) is (see Equation~(\ref{eqn:epsilon_th}))
$\hat{\mathcal{P}}\sim \wc\wD/(\wT^2+\wD^2)$. Thus, our numerical result
shown in Figure~\ref{fig:H_epsilon_w} is consistent with $\wD\sim\wc$ and $\wD$ depends weakly on 
$\wT$ in the $\wT\gg\wc$ regime.

If we interpret the dissipation of the tidal flow found in our simulations
as the result of an effective turbulent viscosity $\nut$, corresponding to a viscous
force term $\nut\nabla^2\dot{\bm\xi}$ in the Navier-Stokes equation,
then our results would imply that $\nut$ does not depend strongly on $\tw$.
We note, however, that the concept of turbulent viscosity may not be appropriate in the $\tw\gg1$
regime, as emphasized by \cite{Terquem2021}.

We also explore the effect of stellar spin on turbulent tidal dissipation
by adding the Coriolis force in our simulations. The result depends on
the rotation frequency $\Omega$ relative to the tidal frequency $\wT$:
(i) slow rotation (with $\Omega\lesssim \wT/2$) has a minor effect on the
tidal power, (ii) moderate rotation (with $\wT/2\lesssim\Omega\lesssim\wT$)
results in the excitation of inertial waves and increases
the tidal power, and (iii) strong rotation (with $\Omega\gtrsim\wT$)
suppresses the tidal power.
Given the simplified geometry of our numerical setup, our simulation does not 
capture the full dynamical effects of inertial waves, e.g. the critical
shear layers produced when the propagation of inertial waves is confined to a
spherical shell \citep{Ogilvie2014}.

\subsection{Discussion}

Our numerical results suggest that tidal dissipation in the stellar
convection zone is not significantly suppressed in the regime
$\wT\gtrsim \wc$.  This is broadly consistent with the
prediction of \cite{Terquem2021}, and inconsistent with the theoretical
arguments that the effective turbulent viscosity is reduced by a
factor of $\wc/\wT$ or $(\wc/\wT)^2$ when
$\wT\gg \wc$.
Our results also disagree with previous numerical simulations on tidal
dissipation in stellar convection zones \citep[e.g.,][]{Duguid+2020a}.
The discrepancy is likely caused by the different setups
of the simulations, e.g., how the equilibrium tidal flow is introduced
and the vertical boundary conditions.  Several conditions of our setup
indeed match those proposed by \cite{TerquemMartin2021}, including the
open vertical boundaries for the convective layer and a radiative top
layer to carry away energy.  Our work implies that these
numerical details can affect the results significantly.

Our results should be considered tentative given that our simulations
are idealized and have several limitations (see also Section~\ref{sec:limitations}): (i) We use an extended
local Cartesian box (size $L$ smaller than but comparable to the
stellar radius) to mimic the stellar convection zone and a periodic
external force to mimic the tidal force; (ii) The scale of the tidal
flow (of order $L$) in our simulations is only 5 times the pressure
scale height $h_0$ of the convective layer; (iii) Our simulations can
only consider $\wT\lesssim 10\wc$, limited by the
numerical resolution of the turbulent flow (or the Reynolds number of
the simulated turbulence); (iv) Most importantly, since our local
setup does not include f-modes, whose existence require global
restoring forces of real stars, our simulation does not give the
correct magnitude for $\xi$ (the equilibrium tidal
displacement) in response to a given tidal force magnitude $F_0\sim
f_0g_0$:
Indeed $\xi_{\rm eq}\sim F_0/(\wT^2
+i\wT\wD)$ in our simulations, while in real stars we have $\xi
\sim F_0/(\wf^2+i\wT\wD)$, and
typically $\wf\gg \wT$.

Applying our inferred scaling $\wD \sim \wc$ and the ``correct''
$\xi$ (with f-mode), we find the tidal power per unit mass in
stellar convection zones to be
\beq
\mathcal{P}\sim \frac{\wc\wT^2 F_0^2}{\wf^4}
\eeq
for the non-rotating cases.
It has been demonstrated by \cite{TerquemMartin2021} that this result can potentially
explain several observation puzzles, including those related to
late-type binaries and hot-Jupiter systems.

During the preparation of this manuscript, we noticed the work of
\cite{Terquem+2026}. We summarize the similarities and differences
between the current work and theirs below.  With slightly different
implementation details, both of the work have (i) employed a free top
surface of the convection region with a cooling function, and (ii)
introduced tidal flows with periodic forcing rather than through a
shearing-box approach.  The major difference between the two works
lies in how tidal dissipation rate is computed: \cite{Terquem+2026}
calculated the transfer term between the mean and the turbulent flows
explicitly, whereas in this work we calculate work done by the tidal
force directly.  Compatible results are obtained, namely that tidal
dissipation is not suppressed as strongly as what is implied in
previous shearing-box setups.

Given the limitations of the local Cartesian setup, extending a similar
convective setup in a spherical wedge seems to be a promising future
step, in which a smaller simulation domain than a full sphere yields
an optimized resolution power for turbulence, yet still allowing for
realistic magnetohydrodynamical processes including inertial waves and
dynamo actions \citep[e.g.,][]{Warnecke+2018}.

\begin{acknowledgments}
We thank Yanqin Wu for helpful discussion.
HZ acknowledges support from the National Natural Science Foundation of China (No. 12403020),
Qimeng Project at Shanghai Polytechnic University,
and the China Postdoctoral Science Foundation (No. 2023M732251).
The numerical simulations in this work were carried out on the Siyuan Mark-I and the ARM platform clusters supported by the Center for High Performance Computing at Shanghai Jiao Tong University,
and the Astro cluster supported by Tsung-Dao Lee Institute.
\end{acknowledgments}

\appendix
\section{Regarding resonant eddy assumption}
\label{appx:w-2}

\cite{Duguid+2020b} studied the wavenumber-space distribution of $\nut$ and found that the energy-dominant eddies always contribute the most, regardless of the tidal frequency.
The results seemingly suggest that the resonant eddies are irrelevant for explaining the $\propto \tw^{-2}$ scaling.
Here we provide an argument based on dimensional analysis that produces the $\tw^{-2}$ scaling without invoking the resonant-eddy assumption.

Let us decompose the turbulent viscosity into the contributions from eddies based on their turnover rate,
\beq
\nut(\wT)=\int \hat{\nu}_\text{T}(\omega) f(\wT,\omega)\ \text{d}\omega,
\label{eqn:nut=fnu}
\eeq
where $\hat{\nu}_\text{T}\propto\omega^{-3}$ is the eddy viscosity per unit frequency at frequency $\omega$,
and $f$ is a dimensionless shape function which describes the relative contributions from each frequency.
For example, within the resonant-eddy assumption,
$f$ would have a compact support near $\omega=\wT$.

In the following we assume a general shape function $f$ that does not necessarily peak at $\wT$.
If $\wT$ falls in the inertial range of the turbulence,
on dimensional grounds,
the shape function should only depend on the ratio between $\wT$ and $\omega$ but not individually on each of them.
Thus we can write
\beq
\nut(\wT)\propto\wT^{-2}\int x^{-3}f(x)\ \text{d}x,\ x\equiv \omega/\wT,
\label{eqn:nut=fnu2}
\eeq
which yields the $\wT^{-2}$ scaling.
Otherwise, if $\wT$ is much smaller than the outer frequency,
$f$ becomes independent of $\wT$.

We note that although $f(x)$ may strongly peak at $x=1$ (i.e., the coupling is the strongest for the resonant eddies), the integrant in Equation~(\ref{eqn:nut=fnu2}) can nevertheless be dominated at $x<1$ (i.e., by large eddies) because of the power-law factor $x^{-3}$.
This resolves the contradiction between the resonant-eddy argument and the spectral decomposition result by \cite{Duguid+2020b}.

\section{Plots of tidal work}
\label{appx:W}

\begin{figure}
\centering
\includegraphics[width=0.9\textwidth]{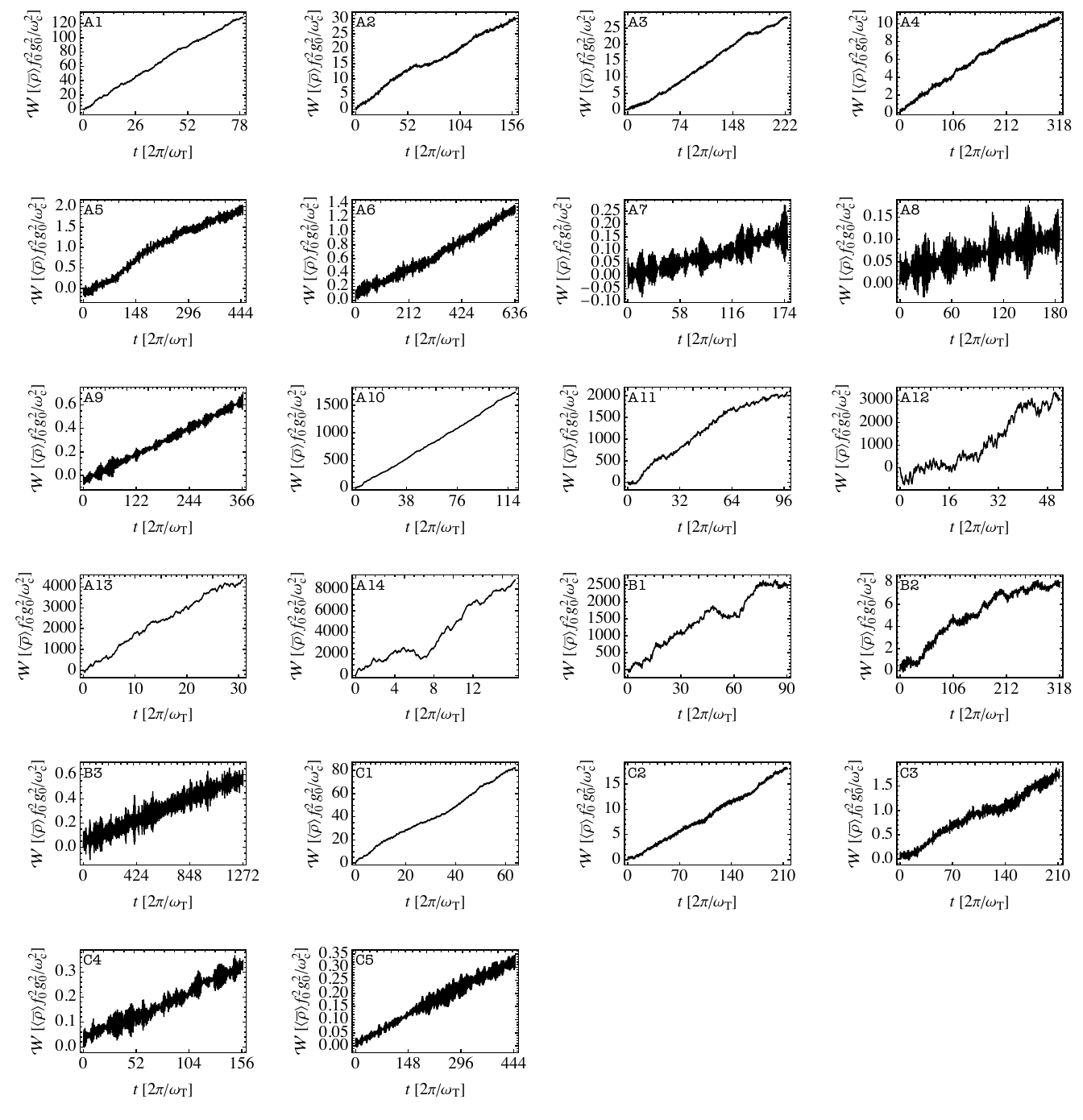}
\caption{The work done by the tidal forcing for the non-rotating runs.}
\label{fig:H_W_vs_t}
\end{figure}

\begin{figure}
\centering
\includegraphics[width=0.9\textwidth]{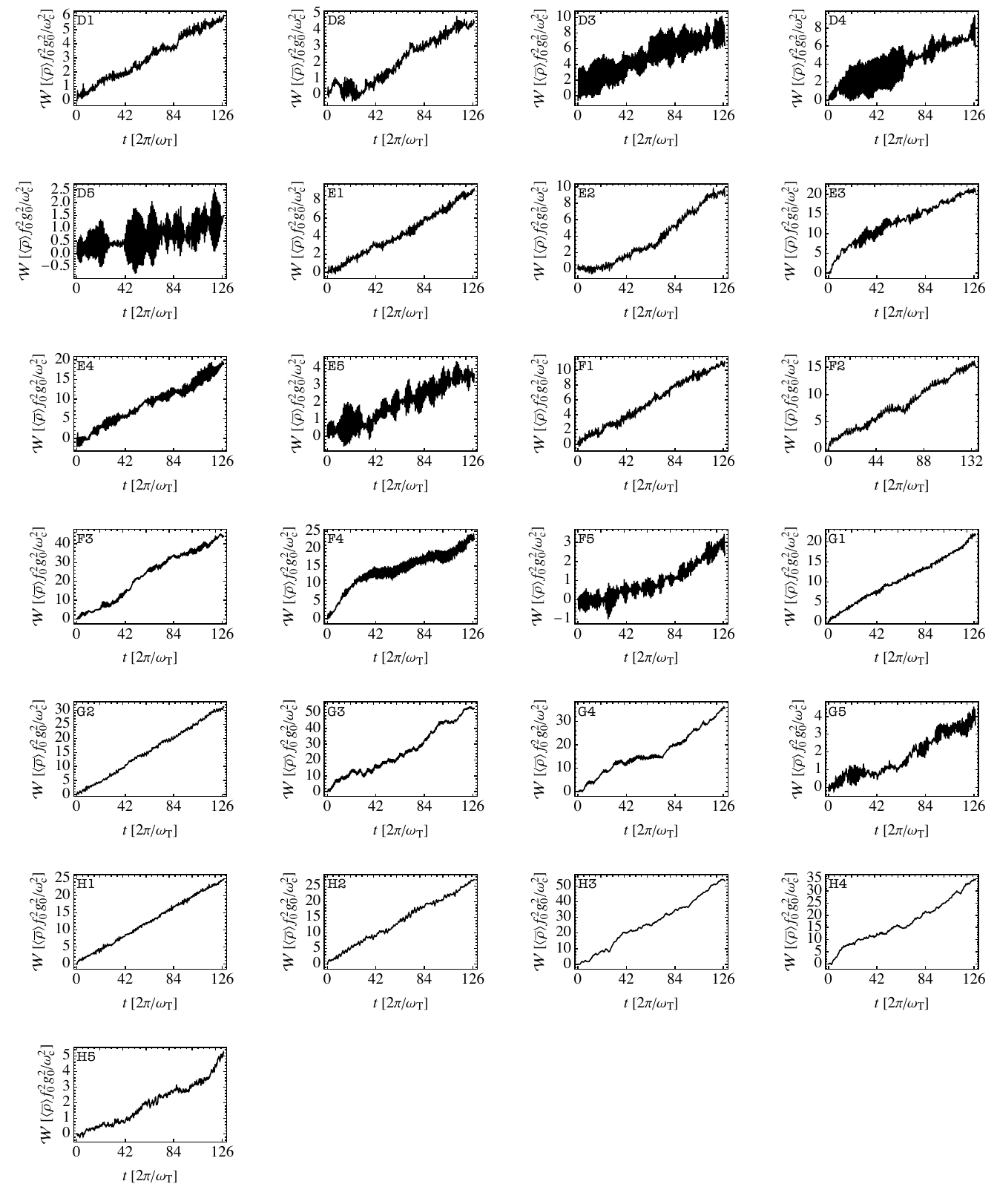}
\caption{The work done by the tidal forcing for the rotating runs.}
\label{fig:HR_W_vs_t}
\end{figure}

Figures~\ref{fig:H_W_vs_t} and \ref{fig:HR_W_vs_t} shows the tidal work done for the non-rotating and the rotating runs, respectively.
All the runs start from a corresponding unforced turbulent steady state (either rotating or non-rotating), but the plotted time has been shifted to start from $0$.

%==============================================

\bibliographystyle{aasjournalv7}
\bibliography{refs}

\end{document}